\documentclass[final,5p,times,twocolumn]{elsarticle}

\usepackage{amssymb}
\usepackage{subcaption} 
\usepackage{gensymb} 
\usepackage{bm}
\usepackage{amsmath}
\usepackage{dblfloatfix} 
\usepackage{booktabs} 
\usepackage{lineno} 

\journal{Nuclear Instruments and Methods in Physics Research, A}

\begin{document}

\begin{frontmatter}



\title{Four-dimensional emittance measurement at the Spallation Neutron Source}


\author{A. Hoover, N. J. Evans} 
\address{Oak Ridge National Laboratory, One Bethel Valley Road, Oak Ridge, Tennessee, 37831, USA}

\begin{abstract}
A coasting hadron beam with an elliptical transverse profile, uniform charge density, and small transverse four-dimensional (4D) emittance could improve accelerator performance in several contexts. A phase space painting method to generate such a distribution is being tested in the Spallation Neutron Source (SNS) accumulator ring. A critical component of these efforts is to measure the 4D emittance of the beam during accumulation. The 4D emittance can be reconstructed from measured beam profiles in two ways: in the multi-optics method, the optics between a reconstruction and measurement location are varied; in the fixed-optics method, multiple measurement locations are used without modifying the optics. The fixed-optics method is faster but can lead to large uncertainty in the reconstructed 4D emittance. In this paper, we implement a variant of the multi-optics method using the four available wire-scanners near the SNS target. We also modify the wire-scanner region to reduce the uncertainty of the fixed-optics method. We then demonstrate the usefulness of the fixed-optics method by reconstructing the 4D emittance evolution during accumulation in the SNS ring.
\end{abstract}







\end{frontmatter}


\section{Introduction}
\label{sec:Introduction}

A coasting hadron beam with an elliptical transverse profile, uniform charge density, and small transverse four-dimensional (4D) emittance could improve accelerator performance in several contexts. First, such a beam could mitigate space charge effects in circular accelerators: the space charge force within the beam would depend linearly on the transverse coordinates, resulting in the minimum possible peak tune shift and zero tune spread \cite{Danilov2003}; at high intensities, a uniform charge density could suppress incoherent space charge resonances due to the lack of higher-order multipoles in the space charge potential; simulations indicate that these resonances can also be mitigated by decreasing the 4D beam emittance \cite{Cheon2022}. Second, a uniform density beam would be ideal for fixed-target applications where the peak density on the target is a concern \cite{Haines2014}. Third, a hadron beam with small 4D emittance could improve collider performance, mainly due to the possibility of rotating the beam in phase space such that the $x$-$y$ projection is completely flat, increasing the luminosity, but also due to the suppression of beam-beam effects \cite{Burov2002, Burov2013}.

Danilov \cite{Danilov2003} proposed a method to generate such a distribution in a ring using phase space painting. Holmes \cite{Holmes2018} computationally investigated the feasibility of the method in the Spallation Neutron Source (SNS) ring. Efforts to implement and test the method in the SNS are ongoing. A critical component of these efforts is to measure the 4D beam emittance during accumulation in the ring.

The covariance matrix of the transverse phase space distribution of the beam, and hence the root-mean-square emittances, can be reconstructed from 1D projections of the distribution onto a horizontal, vertical, and diagonal axis in the transverse plane \cite{book:Minty2003}.\footnote{
    Reconstruction of the transverse phase space distribution is possible using 2D projections of the distribution on a screen \cite{Hock2013, Wang2019, Wolski2020}. This may be possible using the SNS target imaging system \cite{Blokland2010} but is not pursued here.
}At least four measurements are required, each with a different linear transfer matrix connecting the phase space coordinates at the measurement location to the reconstruction location. The transfer matrix can be controlled by varying the machine optics (multi-optics method), the measurement location (fixed-optics method), or any combination of the two \cite{Prat2014}. The SNS has four wire-scanners available along the transfer line connecting the ring to the spallation target. Thus, the fixed-optics method is possible and preferred in our case due to its speed; however, previous studies have shown that the method can be sensitive to errors \cite{Raimondi1993, Woodley2000, Agapov2007, Faus-Golfe2016}. 

The purpose of this work is to optimize the 4D emittance measurement in the SNS. The general reconstruction method is reviewed in Section \ref{sec:Four-dimensional emittance measurement}. The first part of Section \ref{sec:Implementation in the SNS} details the available diagnostics in the SNS; the second part of Section \ref{sec:Implementation in the SNS} describes our implementation of a variant of the multi-optics method in the SNS; the third part of Section \ref{sec:Implementation in the SNS} describes our implementation of the fixed-optics method in the SNS, focusing on reducing the sensitivity to errors. Finally, the usefulness of the fixed-optics method is demonstrated in Section \ref{sec:Example application} by measuring the 4D emittance evolution during accumulation in the SNS ring.

\section{Four-dimensional emittance measurement}
\label{sec:Four-dimensional emittance measurement}

Let $\mathbf{x} = (x, x', y, y')^T$ be the transverse phase space coordinate vector. The covariance matrix $\mathbf{\Sigma}$ is defined by $\mathbf{\Sigma} = \langle{\mathbf{x}\mathbf{x}^T}\rangle$, where $\langle{\dots}\rangle$ represents the average over the distribution; i.e.,
\begin{equation}\label{eq:covariance_matrix}
    \bm{\Sigma} = 
    \begin{bmatrix}
        \langle{xx}\rangle & \langle{xx'}\rangle & \langle{xy}\rangle & \langle{xy'}\rangle \\
        \langle{xx'}\rangle & \langle{x'x'}\rangle & \langle{x'y}\rangle & \langle{x'y'}\rangle \\
        \langle{xy}\rangle & \langle{x'y}\rangle & \langle{yy}\rangle & \langle{yy'}\rangle \\
        \langle{xy'}\rangle & \langle{x'y'}\rangle & \langle{yy'}\rangle & \langle{y'y'}\rangle 
    \end{bmatrix}
    = 
    \begin{bmatrix}
        \bm{\sigma}_{xx} & \bm{\sigma}_{xy} \\
        \bm{\sigma}^T_{xy} & \bm{\sigma}_{yy}
    \end{bmatrix}.
\end{equation}
The covariance matrix defines an ellipsoid in phase space from $\mathbf{x}^T \bm{\Sigma}^{-1} \mathbf{x} = 1$. The 4D emittance $\varepsilon_{4D}$ is proportional to the volume of this ellipsoid and is conserved in any linear focusing system \cite{Lebedev2010}; it is defined by 
\begin{equation}\label{eq:eps4D}
    \varepsilon_{4D} = \left|{\bm{\Sigma}}\right|^{1/2} = \varepsilon_1\varepsilon_2 \le \varepsilon_x\varepsilon_y,
\end{equation}
where $|...|$ is the determinant. We have also introduced the intrinsic emittances $\varepsilon_{1,2}$, which are individually conserved and found by a symplectic diagonalization of $\bm{\Sigma}$ \cite{Dragt2018}; i.e., they are the imaginary components of the eigenvalues of $\bm{\Sigma}\mathbf{U}$ with
\begin{equation}
    \mathbf{U} = 
    \begin{bmatrix}
        0 & 1 & 0 & 0 \\
        -1 & 0 & 0 & 0 \\
        0 & 0 & 0 & 1 \\
        0 & 0 & -1 & 0
    \end{bmatrix}.
\end{equation}
The solution can be written in compact form \cite{Xiao2013}:
\begin{equation}\label{eq:eps12}
    \varepsilon_{1,2} = \frac{1}{2}\sqrt{
      -tr\left[(\bm{\Sigma} \mathbf{U})^2\right] \pm \sqrt{tr^2\left[(\bm{\Sigma} \mathbf{U})^2\right] - 16|{\bm{\Sigma}}|}
    }
\end{equation}
where $tr$ denotes the matrix trace.\footnote{Eq.~\eqref{eq:eps12} does not associate either intrinsic emittance with a specific rotational mode of the beam; instead, $\varepsilon_1$ and $\varepsilon_2$ are ordered by magnitude so that $\varepsilon_1 > \varepsilon_2$, regardless of the sign of the beam's angular momentum (or vorticity, more generally \cite{Groening2021}). We maintain this definition in this work but note that it is inconsistent with the following diagonal form of the covariance matrix:
\begin{equation}
    \mathbf{\Sigma} = 
    \mathbf{V} 
    \begin{bmatrix}
        \varepsilon_1 & 0 & 0 & 0 \\
        0 & \varepsilon_1 & 0 & 0 \\
        0 & 0 & \varepsilon_2 & 0 \\
        0 & 0 & 0 & \varepsilon_2 \\
    \end{bmatrix}
    \mathbf{V}^T
    \nonumber
\end{equation}
where $\mathbf{V}$ is a symplectic matrix and each intrinsic emittance is associated with one mode.\label{fn:1}} The apparent emittances $\varepsilon_x = \left|{\bm\sigma}_{xx}\right|^{1/2}$ and $\varepsilon_y = \left|{\bm\sigma}_{yy}\right|^{1/2}$ are conserved only in uncoupled linear focusing systems and coincide with the intrinsic emittances in the absence of cross-plane correlations ($\bm{\sigma}_{xy} = 0$).

The covariance matrix at position $a$ along the beamline can be reconstructed by measuring $\langle{xx}\rangle$, $\langle{yy}\rangle$ and $\langle{xy}\rangle$ at position $b$, downstream of $a$ \cite{book:Minty2003}. Assuming linear transport, the two covariance matrices are related by $\bm{\Sigma}_b = \mathbf{M} \bm{\Sigma}_a \mathbf{M}^T$, where $\mathbf{M}$ is the linear transfer matrix from $a$ to $b$. Thus, each measurement produces a linear system of equations:
\begin{equation}\label{eq:lsq}
    \begin{bmatrix}
        {\langle{xx}\rangle}^{} \\
        {\langle{xy}\rangle}^{} \\
        {\langle{yy}\rangle}^{} \\
    \end{bmatrix}_b
    = \mathbf{A}
    \begin{bmatrix}
        \langle{xx}\rangle \\
        \langle{xx'}\rangle \\
        \langle{xy}\rangle \\
        \langle{xy'}\rangle \\
        \langle{x'x'}\rangle \\
        \langle{x'y}\rangle \\
        \langle{x'y'}\rangle \\
        \langle{yy}\rangle \\
        \langle{yy'}\rangle \\
        \langle{y'y'}\rangle \\
    \end{bmatrix}_a
    ,
\end{equation}
with \cite{Wolski2020}
\begin{equation}
    \mathbf{A}^T = 
    \begin{bmatrix}
        M_{11}M_{11} & M_{11}M_{31} & M_{31}M_{31} \\
        2M_{11}M_{12} & M_{12}M_{31} + M_{11}M_{32} & 2M_{31}M_{32} \\
        2M_{11}M_{13} & M_{13}M_{31} + M_{11}M_{33} & 2M_{31}M_{33} \\
        2M_{11}M_{14} & M_{14}M_{31} + M_{11}M_{34} & 2M_{31}M_{34} \\
        M_{12}M_{12} & M_{12}M_{32} & M_{32}M_{32} \\
        2M_{12}M_{13} & M_{13}M_{32} + M_{12}M_{33} & 2M_{32}M_{33} \\
        2M_{12}M_{14} & M_{14}M_{32} + M_{12}M_{34} & 2M_{32}M_{34} \\
        M_{13}M_{13} & M_{13}M_{33} & M_{33}M_{33} \\
        2M_{13}M_{14} & M_{14}M_{33} + M_{13}M_{34} & 2M_{33}M_{34} \\
        M_{14}M_{14} & M_{14}M_{34} & M_{34}M_{34}
    \end{bmatrix}
\end{equation}
where $M_{ij}$ is the element in row $i$ and column $j$ of the transfer matrix $\mathbf{M}$. The ten unique elements of $\mathbf{\Sigma}_a$ are obtained by repeating the measurement at least four times with different transfer matrices and computing the linear least squares (LLSQ) solution to the resulting system of equations. The transfer matrix elements can be modified by varying the optics, which we refer to as the multi-optics method, or by varying the measurement location, which we refer to as the fixed-optics method.

\section{Implementation in the SNS}
\label{sec:Implementation in the SNS}

\subsection{SNS facility and diagnostics}

\begin{figure*}[t!]
    \centering
    \begin{subfigure}{0.35\textwidth}
        \includegraphics[width=\textwidth]{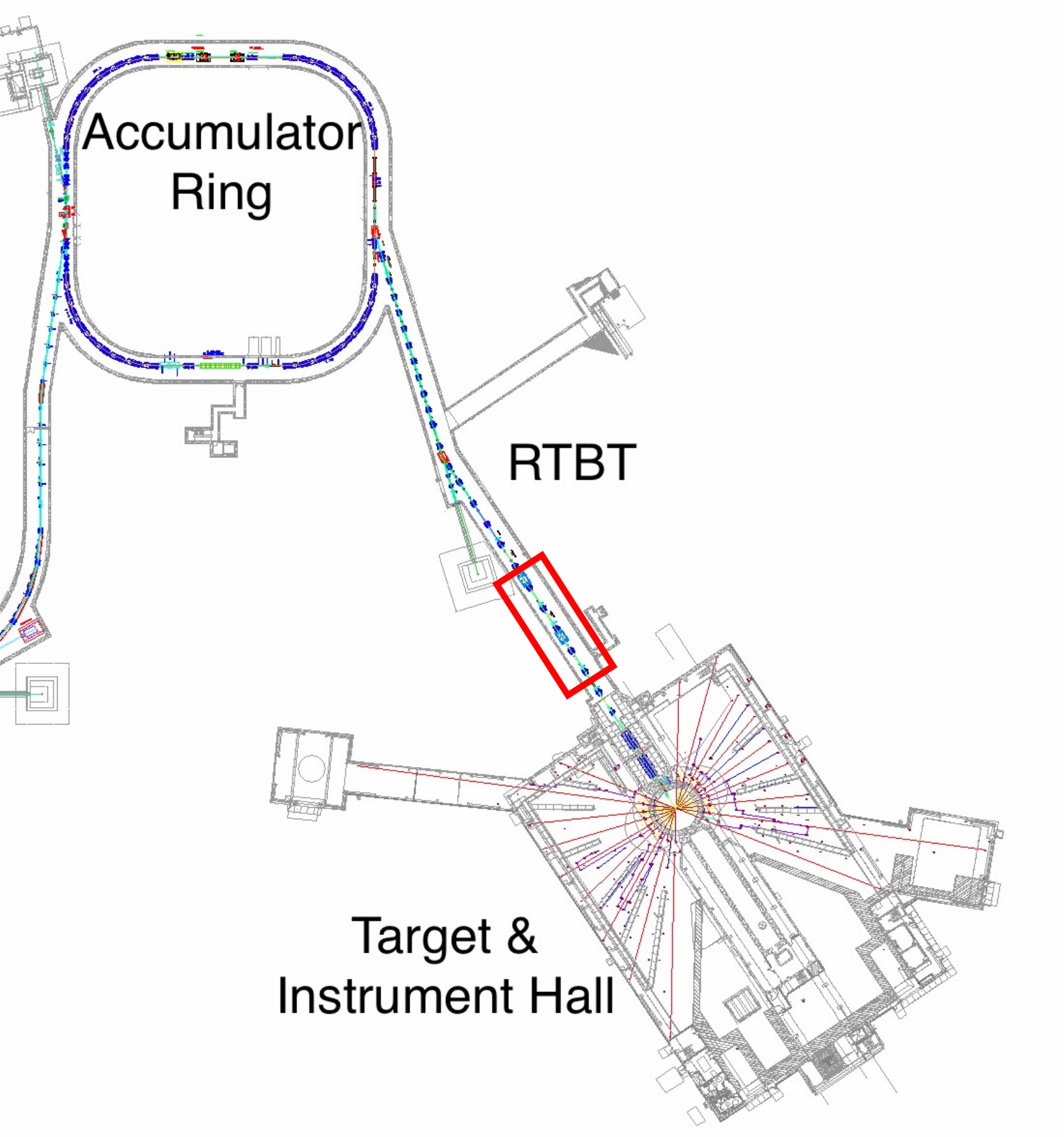}
    \end{subfigure}
    \hspace{0.25cm}
    \begin{subfigure}{0.615\textwidth}
        \includegraphics[width=\textwidth]{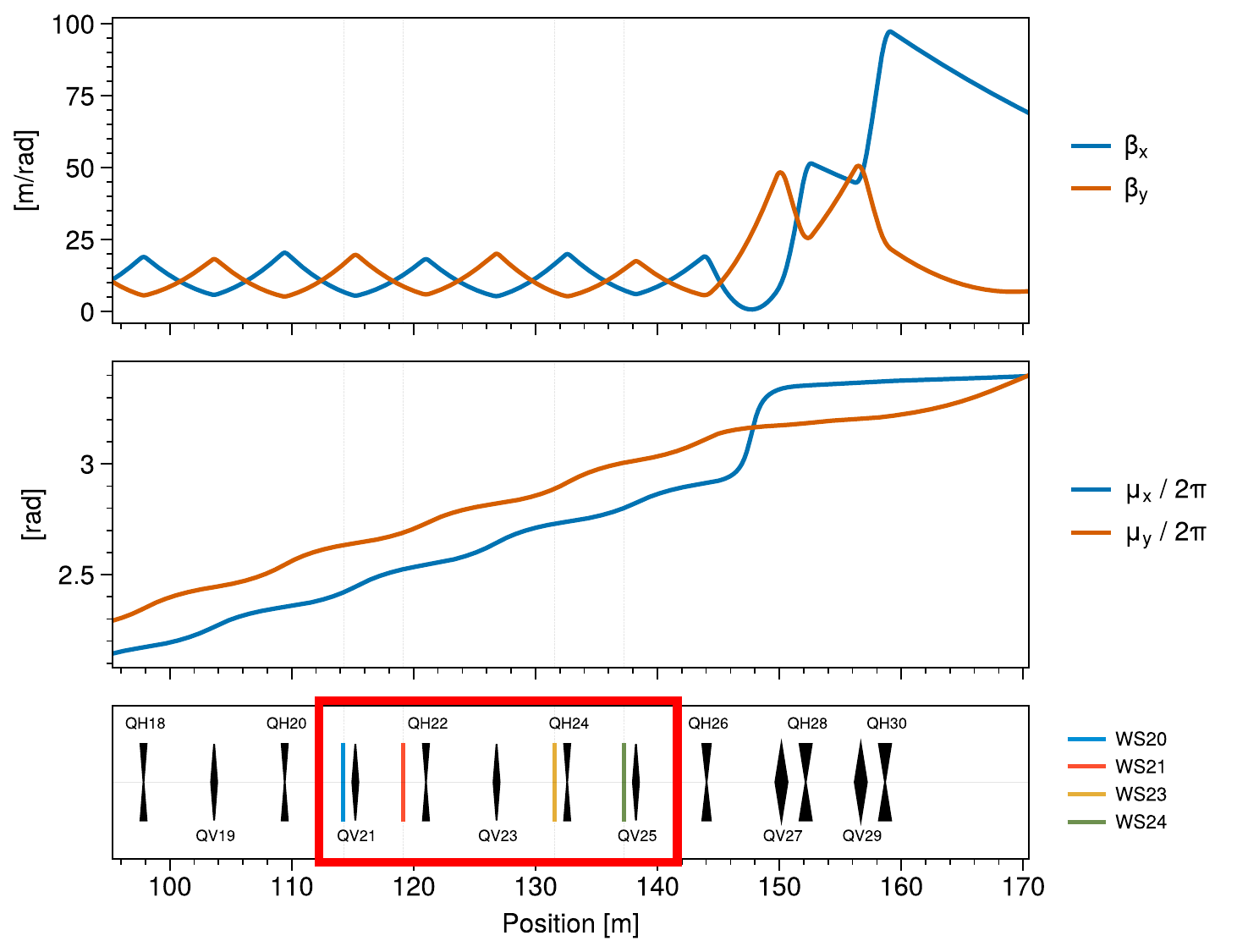}
    \end{subfigure}
    \caption{Left: The accumulator ring and ring-target-beam-transport (RTBT) sections of the Spallation Neutron Source (SNS). Right: $\beta$ functions (top), phase advances (middle), and focusing element and wire-scanner positions (bottom) in the second half of the RTBT. The red box outlines the wire-scanner region.}
    \label{fig:1}
\end{figure*}

The SNS produces short neutron pulses by colliding a proton beam with a liquid mercury target. The beam starts as a continuous wave of H$^-$ ions which is chopped and then bunched in a 402.5 MHz radio-frequency quadrupole (RFQ), forming microsecond-long minipulses. Each minipulse is accelerated in a linac to a kinetic energy of 1 GeV, stripped of its electrons using a carbon foil, and injected into a ring. Approximately one thousand minipulses are accumulated in the ring over $10^{-3}$ seconds. Finally, the proton beam is extracted and guided through the ring-target beam transport line (RTBT) to the target. This process repeats sixty times per second, resulting in an average beam power of 1.4 megawatts. A comprehensive description of the SNS is given in \cite{Henderson2014}; several design parameters of the SNS ring and RTBT are listed in Table~\ref{tab:SNS}; an overhead view of the ring and RTBT is shown in Fig.~\ref{fig:1}.
\begin{table}[!b]
    \centering
    \begin{tabular}{ll}
        \toprule
        \textbf{Quantity} & \textbf{Value} \\
        \midrule
        Ring length & 248 [m] \\
        RTBT length & 151 [m] \\
        Beam kinetic energy & 1 [GeV] \\
        Average beam power & 1.4 [MW] \\
        Repetition rate & 60 [Hz] \\
        Transverse tunes ($\nu_x$, $\nu_y$) & (6.23, 6.20) \\
        Longitudinal tune & $7 \times 10^{-4}$ \\
        \midrule 
    \end{tabular}
    \caption{Several design parameters of the SNS ring and RTBT.}
    \label{tab:SNS}
\end{table}

Diagnostics in the SNS ring are limited to beam position and loss monitors. Beam profile measurements are performed in the RTBT using four wire-scanners, labeled WS20, WS21, WS23, and WS24, the locations of which are shown in Fig.~\ref{fig:1}. Each wire-scanner consists of three 100$\mu m$-thick tungsten wires mounted on a fork at forty-five-degree angles relative to each other. The intensity of secondary electron emission from each wire is recorded as the fork moves across the beam, thus measuring the 1D projection of the beam distribution onto the $x$, $y$, and $u$ axes, where $u$ is tilted 45 degrees above the $x$ axis. The second-order moments $\langle{xx}\rangle$, $\langle{yy}\rangle$, and $\langle{uu}\rangle$ are estimated from the profiles by averaging over the wire positions weighted by the signal amplitude. The cross-plane moment $\langle{xy}\rangle$ is thus measured indirectly: if the tilt angle of the $u$ axis is $\phi$, then $u = x\cos\phi + y\sin\phi$ and 
\begin{equation}
    \langle{xy}\rangle = \frac{\langle{uu}\rangle - \langle{xx}\rangle \cos^2\phi - \langle{yy}\rangle \sin^2\phi}{2\sin\phi\cos\phi}.
\end{equation}
The wire-scanners are run in parallel, taking approximately five minutes to move across the beam and return to their original positions. Their default step size is 3 mm and their dynamic range (the ratio of maximum to minimum measured signal) is approximately $100$.\footnote{The root-mean-square beam sizes may depend on beam tails/halo that cannot be measured by the RTBT wire-scanners. Investigation into the effect of beam halo on emittance measurements in the RTBT is left as future work.} They are run at a beam pulse frequency of 1 Hz.\footnote{Each data point corresponds to a separate beam pulse, so the measurement relies on pulse-to-pulse stability.}

\subsection{Multi-optics method}

We first implemented a variant of the multi-optics method in the SNS. There is considerably more freedom in the choice of optics used in the 4D reconstruction than in 2D reconstruction. We do not consider the general case of coupled optics in this work since the RTBT optics are uncoupled. Thus, the problem reduces to sampling the space of horizontal/vertical phase advances from the reconstruction location to the measurement location such that the reconstruction error is minimized.\footnote{
    In 2D reconstructions, the phase advance from the reconstruction location to the measurement location is typically varied uniformly within a $180\degree$ range. The error is expected to be minimized when the projection angles are spread uniformly between 0 and $\pi$ (the projection angles are only equal to the phase advances in normalized phase space, but it is often convenient to vary the phase advances instead) \cite{Hock2011}. A clear geometric argument is based on the fact that the measured beam sizes define a set of lines that bound an ellipse in 2D phase space at the reconstruction location, so the reconstruction error will be minimized if these lines are evenly distributed around the ellipse. But this argument is less clear in a 4D reconstruction: the measurements of $\left\{\langle{xx}\rangle, \langle{yy}\rangle, \langle{xy}\rangle\right\}$ define a set of surfaces that bound an ellipsoid in 4D phase space at the reconstruction location, and it is not entirely clear how to "evenly distribute" these surfaces around the ellipsoid.
} We chose to follow Prat and Aiba \cite{Prat2014}: in the first half of the scan, the horizontal phase advance is varied while the vertical phase advance is held fixed; in the second half of the scan, the vertical phase advance is varied while the horizontal phase advance is held fixed. Since the four SNS wire-scanners are already spaced somewhat evenly in phase advance, maximal phase coverage is possible by varying the phase advances at each wire-scanner in a $30\degree$ window; however, control of the phase advances between each wire-scanner is limited due to the shared power supplies of the RTBT quadrupoles. There are two power supplies in the wire-scanner region — one controls \{QH18, QH20, QH22, QH24\} and the other controls \{QV19, QV21, QV23, QV25\} — while the last five quadrupoles are controlled independently. In addition to these constraints, the $\beta$ functions must be kept small in the wire-scanner region and must remain close to their nominal values at the target. 

Therefore, we chose to vary the phase advances from the first varied quadrupole (QH18) to the last wire-scanner (WS24), which also modifies the phase advances at the other wire-scanners by similar amounts. To control these phase advances, the two power supplies (eight quadrupoles) in the wire-scanner region are varied to minimize the following cost function:
\begin{equation}
    C(\mathbf{g}) = 
    \left\Vert{
        \bm{\mu} - \tilde{\bm{\mu}}
    }\right\Vert^2
    + 
    \epsilon
    \left\Vert
        \max\left(
            0, \,
            \bm{\beta}_{max} - \tilde{\bm{\beta}}_{max}
        \right)
    \right\Vert^2
    .
\end{equation}
The quadrupole strengths are contained in the vector $\mathbf{g}$; the calculated phase advances are $\bm{\mu} = (\mu_x, \mu_y)$; the desired phase advances are $\tilde{\bm{\mu}} = (\tilde{\mu}_x, \tilde{\mu}_y)$; the maximum calculated $\beta$ functions in the wire-scanner region are $\bm{\beta}_{max} = (\beta_{x_{max}}, \beta_{y_{max}})$; the maximum allowed $\beta$ functions in the wire-scanner region are $\tilde{\bm{\beta}}_{max} = (\tilde{\beta}_{x_{max}}, \tilde{\beta}_{y_{max}})$; $\epsilon$ is a hand-tuned constant. After this process is completed, the remaining five quadrupoles before the target are used to constrain the beam size at the target.

\begin{figure}[b!]
    \centering
    \includegraphics[width=\columnwidth]{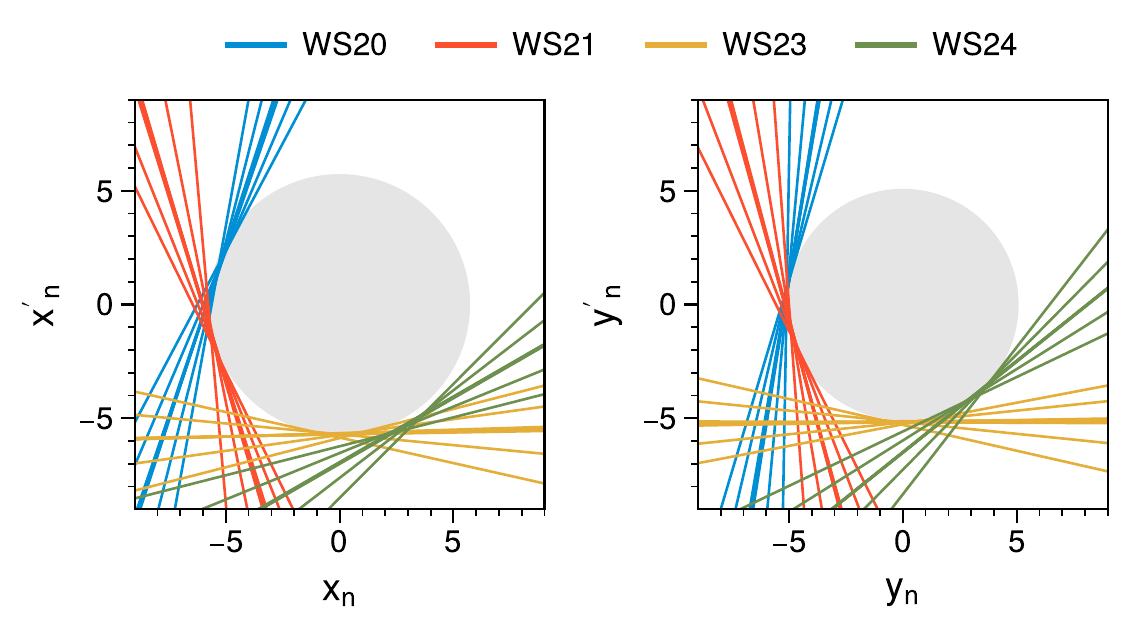}
    \vfill
    \vspace*{0.1cm}
    \vfill
    \begin{tabular}{lll}
        \small\textbf{Parameter} & \small\textbf{Measured} & \small\textbf{Model} \\
        \small$\beta_x$ [m/rad] & \small22.06 $\pm$ 0.29 & \small22.00 \\
        \small$\beta_y$ [m/rad] & \small4.01 $\pm$ 0.02 & \small3.81 \\
        \small$\alpha_x$ & \small2.33 $\pm$ 0.04 & \small2.37 \\
        \small$\alpha_y$ & \small-0.49 $\pm$ 0.01 & \small-0.60 \\
        \small$\varepsilon_1$ [mm~mrad] & \small33.02 $\pm$ \small0.05 & - \\
        \small$\varepsilon_2$ [mm~mrad] & \small25.67 $\pm$ \small1.03 & - \\
        \small$\varepsilon_x$ [mm~mrad] & \small32.85 $\pm$ \small0.05 & - \\
        \small$\varepsilon_y$ [mm~mrad] & \small25.87 $\pm$ \small0.12 & - \\
      \end{tabular}
    \caption{Reconstructed beam parameters and graphical output from a multi-optics emittance measurement of a production beam. Each measurement defines a vertical line at $x = \sqrt{\langle{xx}\rangle}$ and $y = \sqrt{\langle{yy}\rangle}$ in phase space at the measurement location. These lines are transported to the reconstruction location using the transfer matrix and plotted on top of the reconstructed ellipses (gray). The coordinates are normalized by the reconstructed Twiss parameters so that the ellipses are circles with areas proportional to $\varepsilon_{x, y}$.}
    \label{fig:2}
\end{figure}

This multi-optics method was tested on a fully accumulated production beam in the SNS using ten measurements (forty profiles), which were collected in one hour. The reconstructed Twiss parameters, emittances, and best-fit ellipses in the $x$-$x'$ and $y$-$y'$ planes are displayed in Fig.~\ref{fig:2}.
The uncertainties in the parameters were calculated by propagating the standard deviations of the ten reconstructed moments obtained from the LLSQ estimator \cite{Faus-Golfe2016}. The reconstructed Twiss parameters are close to the model parameters computed from the linear transfer matrices of the ring and RTBT, showing that the beam was matched to the nominal optics when it was extracted from the ring. The intrinsic emittances are almost equal to the apparent emittances, showing that there was very little cross-plane correlation in the beam. This is expected for a production beam.

\begin{figure}
    \centering
    \includegraphics[width=0.5\textwidth]{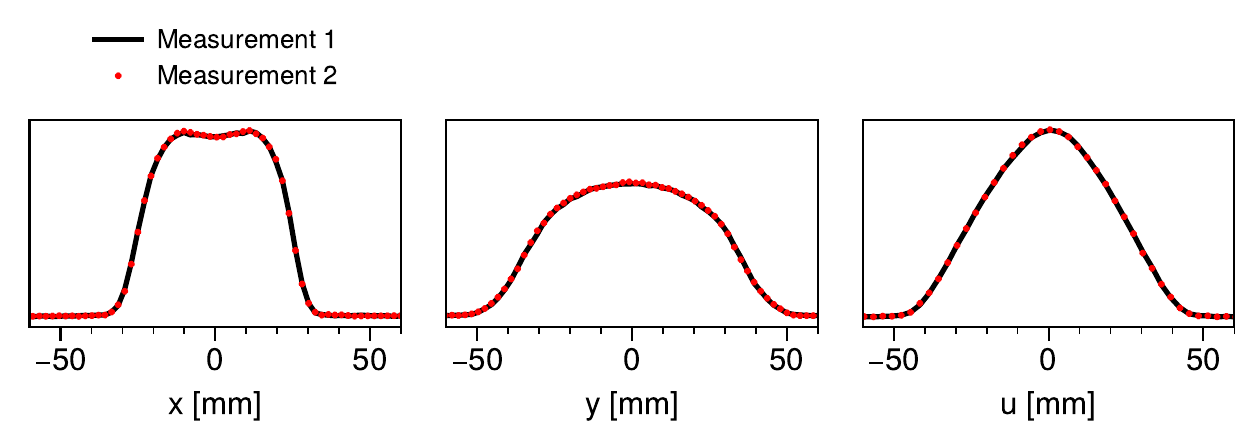}
    \caption{Measured profiles of a production beam along the horizontal ($x$), vertical ($y$), and diagonal ($u$) axis at wire-scanner WS20 in the RTBT. Two measurements are overlayed with identical machine settings.}
    \label{fig:3}
\end{figure}

A comprehensive study of errors in the multi-optics 4D emittance measurement was completed at the SwissFEL Injector Test Facility (SITF) by Prat and Aiba in \cite{Prat2014}. They considered errors in the measured moments, quadrupole field and alignment errors, beam energy errors, beam mismatch at the reconstruction point, and dispersion/chromaticity \cite{Mostacci2012}, concluding that the method remained accurate and reporting $< 5\%$ uncertainty in the intrinsic emittances. We initially performed similar studies using envelope tracking to estimate the reconstruction errors in the RTBT \cite{Hoover2021-IPAC}. We considered uncertainty in the beam moments ($\pm 5\%$) quadrupole tilt angles ($\pm\%$ 1 mrad), quadrupole field strengths ($\pm$ 1\%), beam kinetic energy ($\pm$ 3 MeV for a 1 GeV beam), and beam Twiss parameters at the reconstruction point ($\pm$ 5\%), with all other effects turned off. The emittances were successfully reconstructed with systematic and random errors of a few percent. Most of these uncertainties had only a small effect on the reconstructed intrinsic emittances, with the largest coming from beam mismatch and errors in the measured beam size. Therefore, we only consider these sources of uncertainty going forward.\footnote{Space charge forces, which can render the method invalid for high-perveance beams \cite{Anderson2002}, can be neglected since the space charge tune shift in the ring is around 3\% and the distance from the reconstruction location to the measurement location is much less than the length of the ring.} From repeated wire-scanner measurements with the same machine setup (see Fig.~\ref{fig:3}), we estimated the error in the measured moments to be only 3\%. This 3\% noise will be assumed throughout the rest of the analysis.

\subsection{Fixed-optics method}

The downside of the multi-optics method is its long execution time, for the following reasons. First, we are not only interested in the beam emittances at a single time but are also interested in the emittance evolution throughout accumulation, which conveys valuable information about the beam dynamics and allows qualitative comparison with computer simulation. Second, it would be beneficial to quickly evaluate various machine states: although computer simulation can be used as a guide in choosing the experimental parameters, online tuning is expected to be necessary since small changes to the accelerator may have large effects on the painted phase space distribution. Third, the time reserved for accelerator physics experiments at the SNS is limited, and our initial experiments will require substantial setup time due to modification of the beam energy. Thus, the fixed-optics method is preferred. A modest reduction in accuracy for the increase in speed is warranted since weak cross-plane correlations ($\varepsilon_1 \varepsilon_2 \approx \varepsilon_x \varepsilon_y$) are uninteresting for our purposes and do not need to be resolved. 

Unfortunately, the nominal optics in the RTBT are ill-suited for the fixed-optics reconstruction: if only one set of optics from Fig.~\ref{fig:2} is used, the resulting covariance matrix is not positive-definite. We label this a failed fit. A nonlinear solver \cite{Raimondi1993} or Cholesky decomposition \cite{Agapov2007} can be used to ensure a valid covariance matrix, but we found that the answer depended strongly on the initial guess provided to the solver and on which measurement in the scan was used in the reconstruction.

\begin{figure}[!b]
    \centering
    \includegraphics[width=\columnwidth]{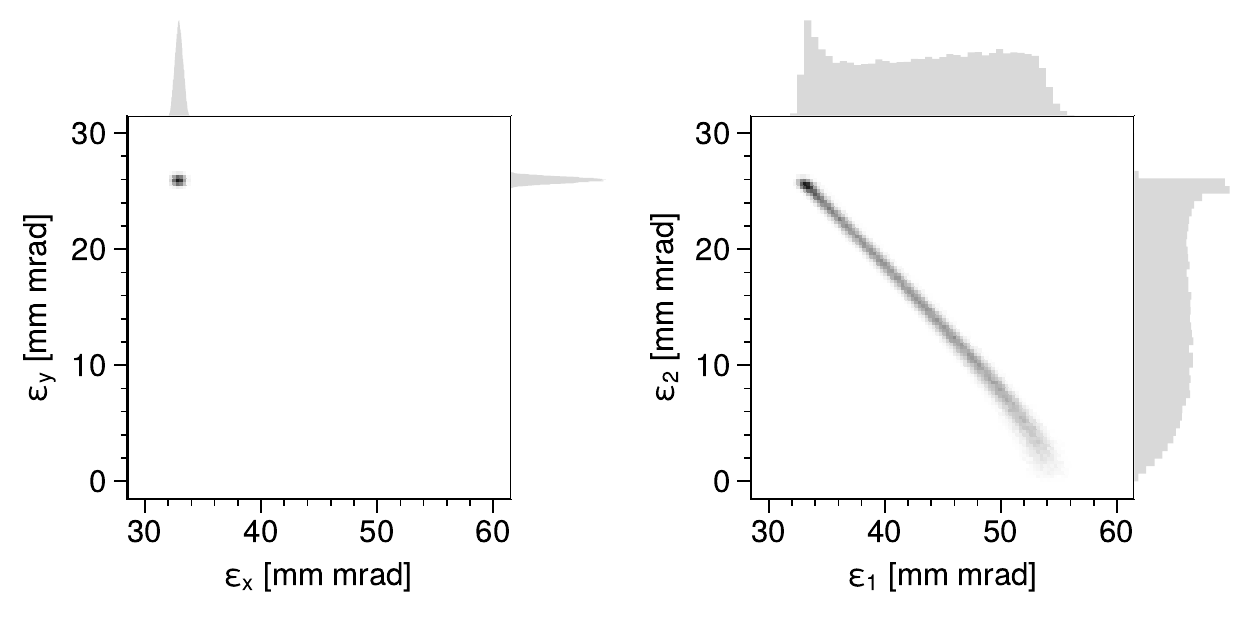}
    \caption{Simulated emittance measurement using the nominal RTBT optics. The true values are $\varepsilon_1 = \varepsilon_x$ = 32 mm~mrad and $\varepsilon_2 = \varepsilon_y$ = 25 mm~mrad.}
    \label{fig:4}
\end{figure}

To investigate the sensitivity of the fixed-optics method, we generated a covariance matrix with $\varepsilon_1 = \varepsilon_x$ = 32 mm~mrad and $\varepsilon_2 = \varepsilon_y$ = 25 mm~mrad, close to the measured values, and with Twiss parameters matched to the lattice. This covariance matrix was tracked to the wire-scanners using the transfer matrices from the fifth step in the scan of Fig.~\ref{fig:2}, and the reconstruction was performed many times with 3\% random noise added to the $\langle{xx}\rangle$, $\langle{yy}\rangle$, and $\langle{uu}\rangle$ moments. (We ignored any uncertainty in the transfer matrix elements, which were expected to be small.) This resulted in a large fraction of failed trials, but some successful trials. Fig.~\ref{fig:4} shows the emittances in the successful trials.
Unlike the apparent emittances, the intrinsic emittances are strongly correlated and are not centered on the correct values. Our primary concern is to reduce the latter effect, which we refer to as bias.

Sensitivity of fixed-optics 4D emittance measurements was observed by Woodley and Emma \cite{Woodley2000} and studied more recently by Agapov, Blair, and Woodley \cite{Agapov2007} as well as Faus-Golfe et al. \cite{Faus-Golfe2016}, all in the context of design studies for a future International Linear Collider (ILC). The motivation for these studies was to remove the cross-plane correlation in the beam to minimize the vertical emittance $\varepsilon_y$. Woodley and Emma proposed to abandon the fixed-optics method due to the bias in the intrinsic emittances introduced by large errors in the measured moments, suggesting to instead measure the apparent emittances and iteratively minimize $\varepsilon_y$. Agapov, Blair, and Woodley revisited this problem and showed that the linear system used to reconstruct the cross-plane moments can easily become ill-conditioned, and hence very sensitive to errors in the measured moments, suggesting the use of the condition number $C = |\mathbf{A}| |\mathbf{A}^{-1} |$, where $| \dots |$ is a matrix norm \cite{Golub1985} and $\mathbf{A}$ is defined in Eq.~\eqref{eq:lsq}, to characterize the sensitivity. Faus-Golfe et al. built on this work, studying the problem analytically. They suggested that the optics in the planned ILC emittance measurement station, which contained four wire-scanners, could be modified to allow the use of the fixed-optics reconstruction.

\begin{figure}[]
    \centering
    \includegraphics[width=\columnwidth]{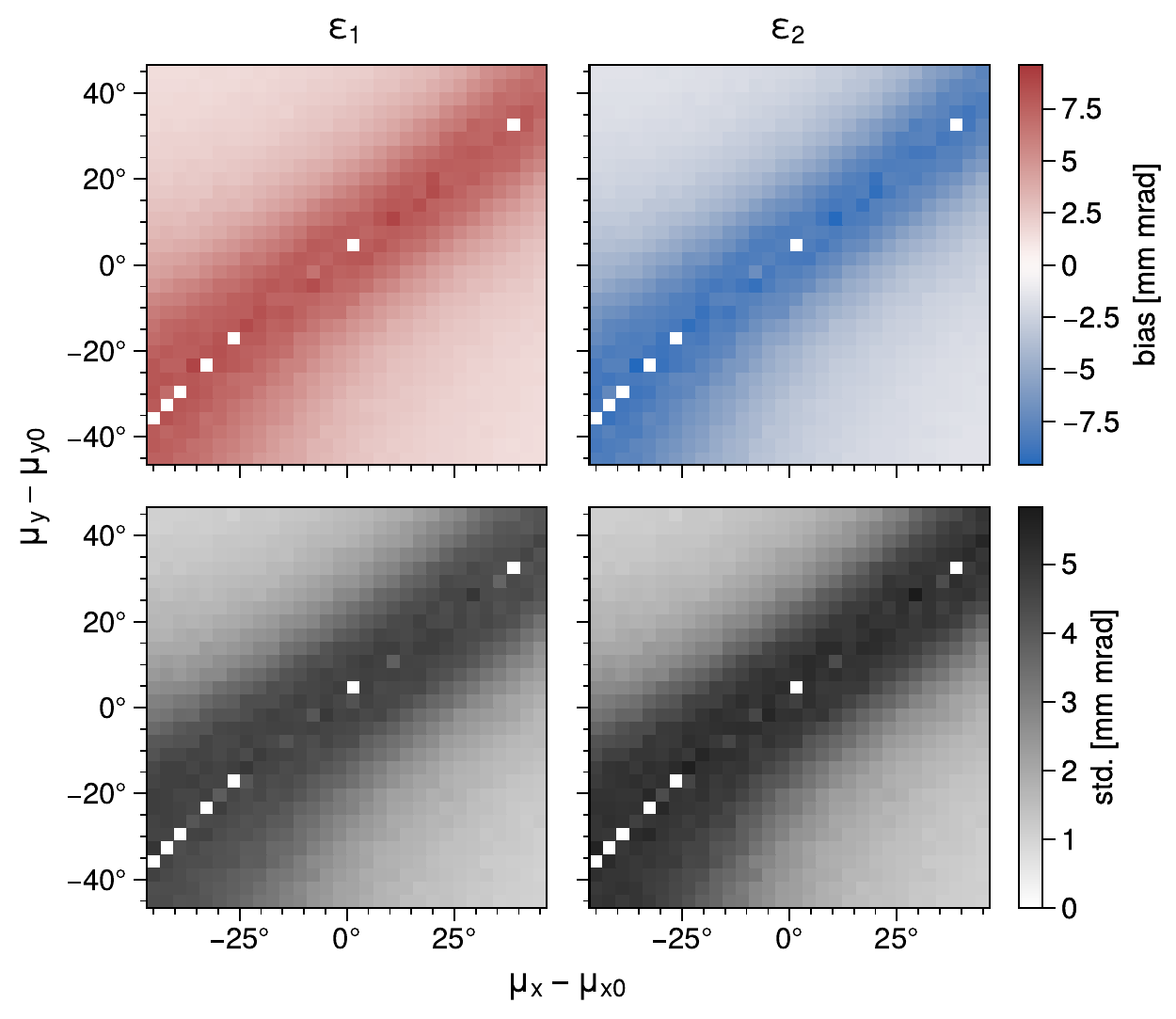}
    \caption{Simulated bias and standard deviation of the intrinsic emittances in the RTBT as a function of the phase advances at WS24. The true values are $\varepsilon_1 = \varepsilon_2 = \varepsilon_x = \varepsilon_y$ = 20 mm~mrad. Settings that produced no successful trials appear as white cells.}
    \label{fig:5}
\end{figure}

We performed a similar modification of the RTBT optics, using these previous studies as a guide. Recall the two available knobs: the two power supplies in the wire-scanner region used to control the phase advances at the final wire-scanner. To search for a better set of optics, we varied these phase advances ($\mu_x$, $\mu_y$) in a $90\degree$ window around their nominal values ($\mu_{x0}$, $\mu_{y0}$). At each setting, a matched covariance matrix with $\varepsilon_1 = \varepsilon_2 = \varepsilon_x = \varepsilon_y$ = 20 mm~mrad was generated and tracked to the wire-scanner locations. The reconstruction was again simulated with 3\% random noise added to the ``measured" moments, repeating over a few thousand trials. The mean and standard deviation of the emittances were calculated over the successful trials. The difference between the mean emittances and the true emittances, which we refer to as the \textit{bias}, is plotted for each set of optics in the top row of Fig.~\ref{fig:5}, as well as the standard deviations in the bottom row.
\begin{figure*}[t!]
    \centering
    \begin{subfigure}[b]{0.92\textwidth}
       \includegraphics[width=\linewidth]{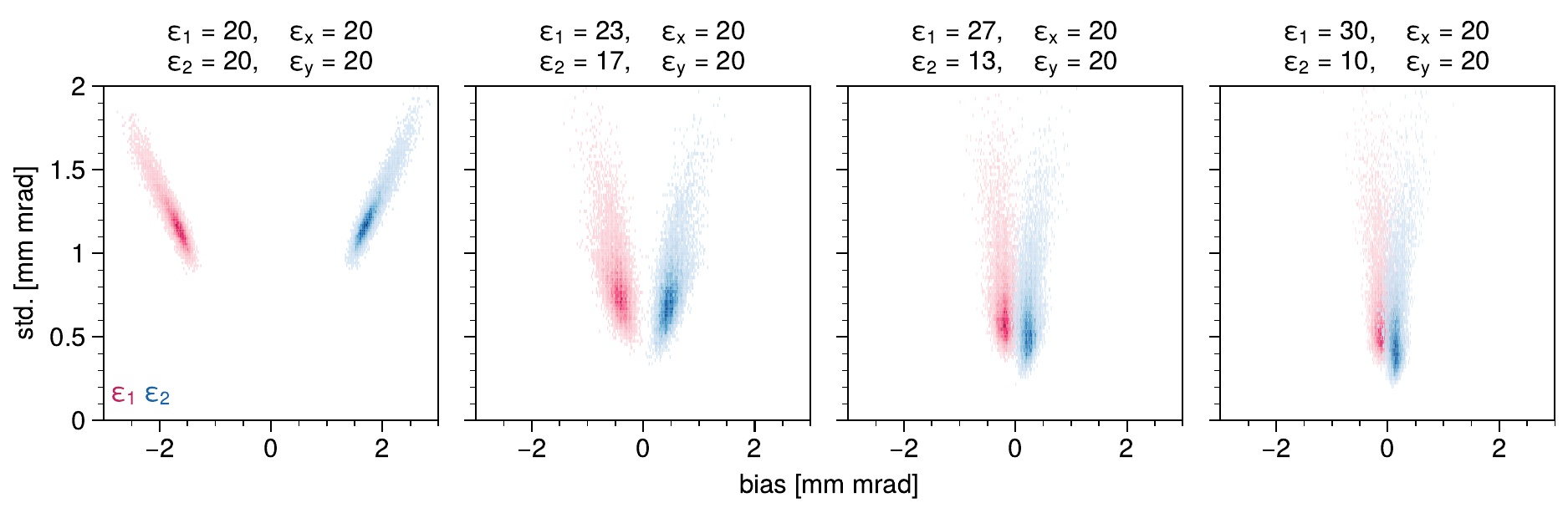}
       \caption{}
       \label{fig:6a} 
    \end{subfigure}
    \begin{subfigure}[b]{0.92\textwidth}
       \includegraphics[width=\linewidth]{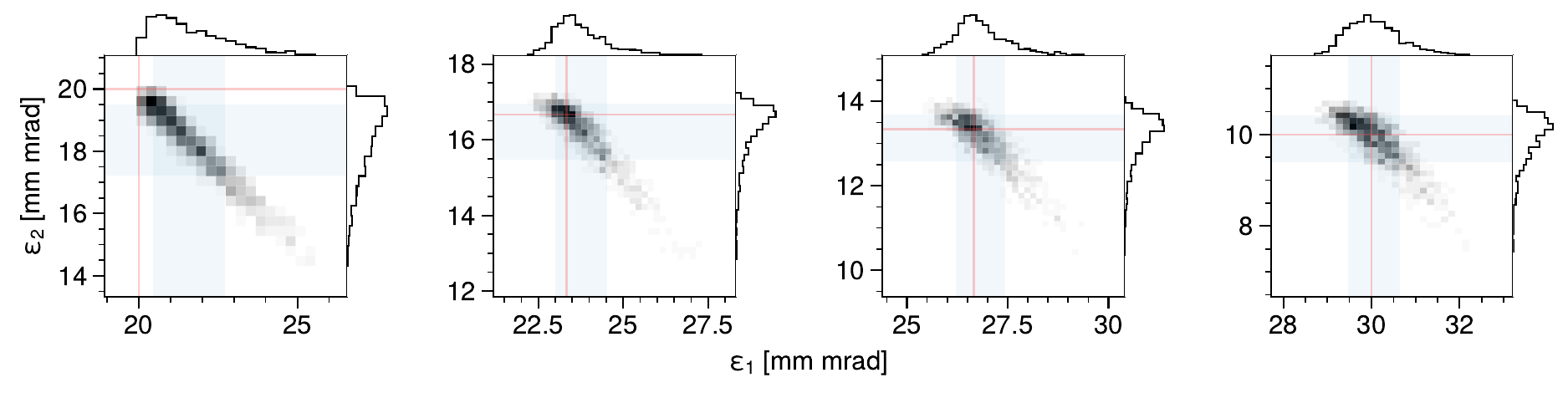}
       \caption{}
       \label{fig:6b}
    \end{subfigure}
    \caption{(a) Bias and standard deviations of $\varepsilon_1$ (pink) and $\varepsilon_2$ (blue) from simulated reconstructions in the RTBT as the beam Twiss parameters $\left\{\alpha_x, \alpha_y, \beta_x, \beta_y\right\}$ are scanned within a four-dimensional grid around their nominal values (as defined by the lattice). Each point in each subfigure corresponds to a simulation with different Twiss parameters. The beam emittances are fixed in each column, varying from no cross-plane correlation on the far left to large cross-plane correlation on the far right.  (b) For each subfigure in (a), from left to right, we plot the distribution of intrinsic emittances generated by the random trials for one set of Twiss parameters (greyscale). Each blue shaded band is centered at the mean of the distribution along one axis, with a width equal to twice the standard deviation along that axis. The red lines show the true values of the emittances.}
    \label{fig:6}
\end{figure*}
The apparent emittances are not displayed because they remained within 1\% of their true values at every optics setting. Modifying the optics so that $\mu_x - \mu_{x0} = 45\degree$ and $\mu_y - \mu_{y0} = -45\degree$ reduces the bias to $\approx 7\%$ and the standard deviation to $\approx 5\%$. The fraction of failed fits, which is very large along the diagonal in the figure, is reduced to zero at this setting.

It is also important to examine the effect of mismatched beam parameters — $\alpha_x$, $\beta_x$, $\alpha_y$, $\beta_y$ — on the accuracy of the reconstructed intrinsic emittances, just as was done for the apparent emittances in \cite{Prat2014}. All previous phase advance calculations have assumed that the beam Twiss parameters are the same as the ring Twiss parameters at extraction. It is possible, however, for space charge to effectively modify the ring Twiss parameters, resulting in mismatch when entering the RTBT. This modification is small for the normal injection scheme, as shown in Fig.~\ref{fig:2}, but we have generated more significant mismatch in our studies of non-standard injection schemes at lower beam energies. One example is shown in Table~\ref{tab:mismatch}.
\begin{table}[!b]
    \centering
    \begin{tabular}{lll}
        \toprule
        \textbf{Parameter} & \textbf{Measured} & \textbf{Model} \\
        \midrule
        $\beta_x$ [m] & 6.26 & 5.49 \\
        $\beta_y$ [m] & 20.82 & 19.25 \\
        $\alpha_x$ & -0.89 & -0.78 \\
        $\alpha_y$ & 1.17 & 1.91 \\
        \midrule 
    \end{tabular}
    \caption{Measured and model beam Twiss parameters in the RTBT.}
    \label{tab:mismatch}
\end{table}

The beam mismatch is unlikely to exceed these values in our studies. To examine the effect of mismatch, we first moved the operating point to $\mu_x = \mu_{x0} + 45\degree$, $\mu_y = \mu_{y0} - 45\degree$, then varied $\beta_x$ and $\beta_y$ within a $\pm 20\%$ window around their model values, $\alpha_x$ within a $\pm 15\%$ window, and $\alpha_y$ within a $-40\%, +10\%$ window to extend beyond the measured discrepancies, and repeated the Monte Carlo trials for each initial beam, producing a collection of means and standard deviations for the reconstructed intrinsic emittances. The left plot in Fig.~\ref{fig:6a} displays the standard deviations and biases for $\varepsilon_1$ (pink) and $\varepsilon_2$ (blue).

Although most of the points are clustered near the original bias and standard deviation of 7\% and 5\%, respectively, the bias increases to nearly 15\% in some cases, which may make it difficult to resolve weak cross-plane correlation; however, the measurement should still resolve strong cross-plane correlation. This is demonstrated in the rest of the plots in Fig.~\ref{fig:6}a, in which the entire process is repeated with $\varepsilon_1 / \varepsilon_2 > 1$. The bias in the reconstruction quickly decreases — the emittances are clustered around their true values. We conclude that with modifications to the RTBT optics, the fixed-optics method should be sufficient for faster (albeit less accurate) four-dimensional emittance measurements in the SNS.

(The explanation for the dependence of the bias on the cross-plane correlation in the beam is the following: The perturbations applied to the covariance matrix in our random trials produce only small changes to the reconstructed apparent emittances. If the reconstructed apparent emittances are assumed to be constant over the trials, the 4D emittance is bounded from above by $\varepsilon_1\varepsilon_2 \le \varepsilon_x\varepsilon_y$. If there is no cross-plane correlation in the beam so that $\varepsilon_1\varepsilon_2 = \varepsilon_x\varepsilon_y$ — as is the case in the left column of Fig.~\ref{fig:6a} — the mean of the distribution of 4D emittances will be less than the true value since $\varepsilon_1\varepsilon_2$ is always less than or equal to $\varepsilon_x\varepsilon_y$. If there is significant cross-plane correlation in the beam — as is the case in the right column of Fig.~\ref{fig:6a} — the 4D emittance has more freedom to increase or decrease without violating Eq.~\eqref{eq:eps4D} and may, therefore, be centered on the correct value. As discussed in Footnote~\ref{fn:1}, Eq.~\eqref{eq:eps12} orders the intrinsic emittances by magnitude; as a consequence of this ordering, as well as the constraint just discussed, $\varepsilon_1$ is bounded from below and $\varepsilon_2$ is bounded from above. This is evident in Fig.~\ref{fig:6b}, which plots the distribution of intrinsic emittances for one set of Twiss parameters from each plot in Fig.~\ref{fig:6a}.)

\section{Example application}
\label{sec:Example application}

We now demonstrate the use of the fixed-optics method to measure the intrinsic and apparent emittances of a beam during accumulation in the SNS ring. The purpose of this section is not to analyze the measurement results in any detail; rather, it is to demonstrate that the measurement is responsive to changes in the ring lattice and injection parameters.

It is first necessary to describe our approach to assigning uncertainty, in the form of error bars, to the emittances reconstructed by the fixed optics method. Propagating the uncertainties in the reconstructed moments, as obtained from the LLSQ estimator, is not an option since no least-squares fit is performed — the cross-plane moments are exactly determined by the four measured profiles. Since the wire-scanners take over five minutes to measure the beam profiles, it is impossible to compute the mean and standard deviation over many trials. We propose to use a similar approach as in the previous section: repeatedly perturb the measured moments in a 3\% range, each time reconstructing the covariance matrix and computing the emittances; compute the mean and standard deviation of the emittances over all trials; plot an error bar centered at the mean, with a width equal to twice the standard deviation. 

\begin{figure}[t!]
    \centering
    \includegraphics[width=\columnwidth]{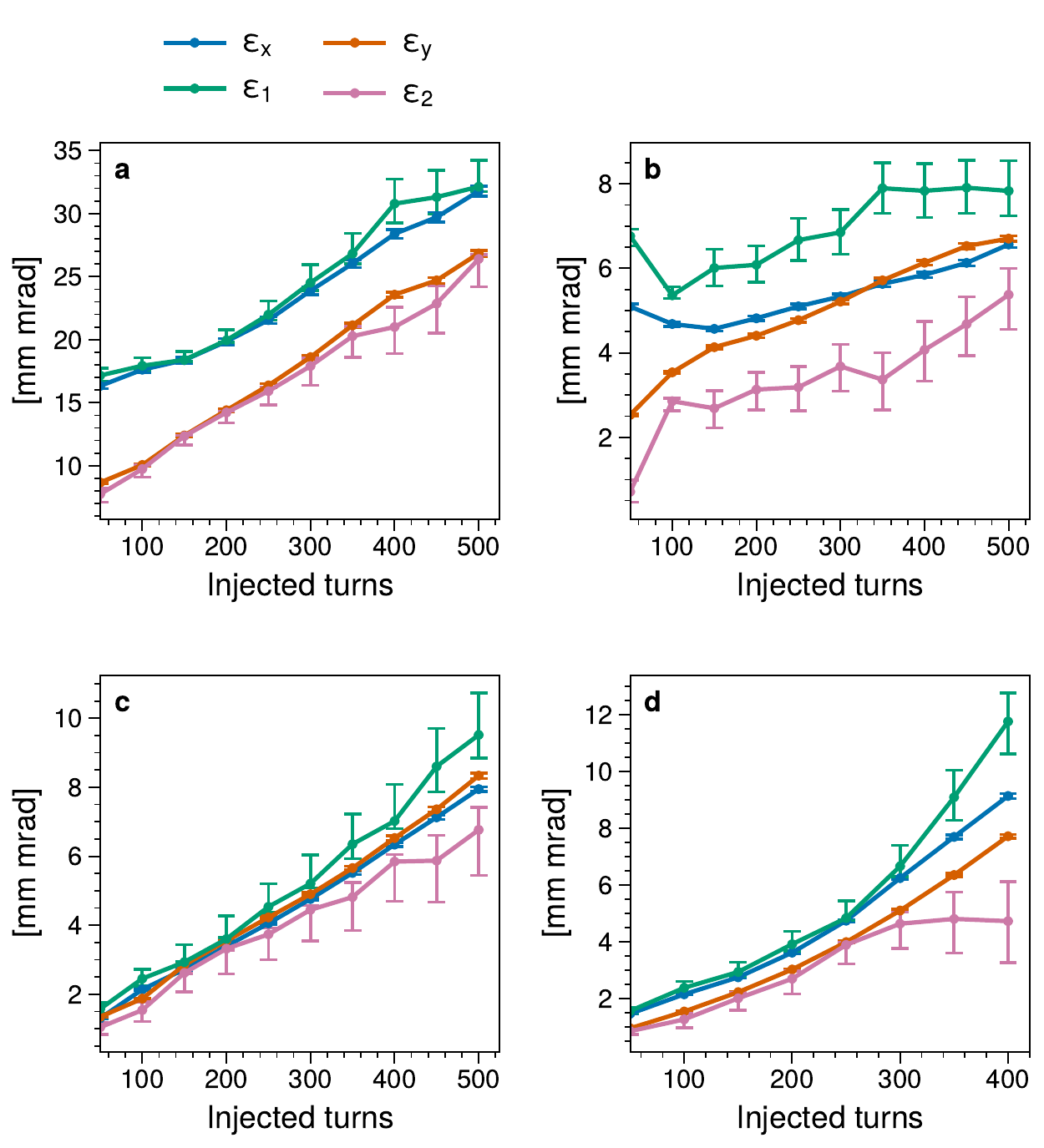}
    \caption{Measured emittances vs. number of injected turns in the SNS ring. (a) correlated (production) painting at 1 GeV; (b) attempted elliptical painting at 1 GeV; (c)-(d) elliptical painting at 0.8 GeV. In (d), the painted beam size was increased and the beam intensity was decreased relative to (c). Each error bar is calculated from the mean and standard deviation of 5000 trials: in each trial, the emittances are reconstructed from a random 3\% perturbation of the measured moments.}
    \label{fig:7}
\end{figure}

We first display the measured emittance evolution of a 1 GeV beam produced by the standard SNS painting method.\footnote{The standard painting method is often referred to as "correlated painting". The distance between the injected and circulating beam is varied as the square root of time while the angle between the beams is fixed at zero. Note that ``correlated painting" tends not to produce any cross-plane correlation in the beam. See \cite{Wang1999, Hotchi2020}.} The beam was accumulated over 500 injected turns; it was extracted and measured every 50 turns during accumulation. The reconstructed emittances throughout accumulation are shown in Fig.~\ref{fig:7}a.
As expected, the beam did not develop any significant cross-plane correlation. The beam Twiss parameters were essentially matched at the nominal optics at the RTBT entrance, so bias in the measurement was not a concern.

The rest of the figures repeat this measurement using a different painting method, which we now briefly describe. As mentioned in Section 1, our goal is to produce a beam with a uniform charge density, elliptical transverse profile, and small 4D emittance. A proposed method to generate such a distribution is called ``elliptical painting" \cite{Holmes2018}: the working principle is to generate elliptical modes in the ring — elliptical turn-by-turn trajectories of particles in the $x$-$y$ plane — and to inject particles into one of the modes, scaling the particle amplitudes with square root time-dependence.\footnote{The turn-by-turn trajectory of a particle in the $x$-$y$ plane is an ellipse if its phase space coordinates lie along an eigenvector of the ring transfer matrix. If there are no coupled elements in the ring, the eigenvectors correspond to uncoupled horizontal or vertical motion. In general, coupled elements must be used to create elliptical trajectories. However, there is a special case in which the $x$ and $y$ tunes are equal; in this case, any vector is an eigenvector, so any initial phase space coordinates will result in elliptical modes. See \cite{Hoover2022-thesis}.} If the method works as intended, a uniform density ellipse with near-zero 4D emittance is maintained throughout accumulation. The key difference from the standard SNS painting scheme is that the angles between the circulating and injected beams must be changed during injection, not just the distances. We denote the distances as $x$ and $y$ and the angles as $x'$ and $y'$ in this section. The most promising painting path in the SNS is a line in the $x$-$y'$ plane with a slope chosen such that the apparent emittances are equal, resulting in an approximately circular distribution at the injection point. These injection parameters are measured and controlled using beam-position-monitor (BPM) waveforms in the ring and eight time-dependent injection kicker magnets.

Figure Fig.~\ref{fig:7}b represents an initial attempt to perform elliptical painting in the SNS ring. The conditions were not ideal: First, due to hardware constraints, the required initial injection coordinates $x = x' = y = y' = 0$ could not be obtained at the nominal beam energy of 1 GeV; the closest possible coordinates were ($x$, $x'$, $y$, $y'$) = (10 mm, 0 mrad, 0 mm, 0 mrad), meaning that the initial distribution would have a hollow center and would not have a uniform charge density. Second, the maximum $y'$ was quite small at 0.7 mrad, placing a limit on the vertical beam size. Third, because no coupled elements existed in the SNS ring, elliptical modes were generated by equating the fractional parts of the horizontal and vertical tunes; it was expected that such a setup would be sensitive to the tune split and that nonlinearities would strongly influence the beam dynamics near the difference resonance $\nu_x \approx \nu_y$. Therefore, we did not expect the painted beam to approach the ideal case of zero 4D emittance. However, we note that the emittance evolution is significantly different than in Fig.~\ref{fig:7} and that the intrinsic emittances are not the same as the apparent emittances. 

Fig.~\ref{fig:7}c shows a second experiment in which the beam energy was lowered to 0.8 GeV, allowing initial injected coordinates $x \approx x' \approx y \approx y' \approx 0$, as required by the painting method, and final injection coordinates ($x$, $x'$, $y$, $y'$) $\approx$ (21 mm, 0 mrad, 0 mm, 1.1 mrad). Simulations predicted that a small cross-plane correlation would develop in the beam after a few hundred turns, and although this was measured, it is on the order of the expected bias.\footnote{The beam Twiss parameters were more significantly mismatched at the RTBT entrance in this experiment, likely due to the increased space charge intensity, leading to larger uncertainties in the intrinsic emittances. The measured Twiss parameters are listed in Table \ref{tab:mismatch}.} In Fig.~\ref{fig:7}d, the measurement was repeated for a larger beam size and lower beam intensity, which simulations predicted would reduce the 4D emittance during accumulation. In this case, the differences between the intrinsic and apparent emittances are larger than the expected bias in the reconstruction. 

\begin{figure}[t!]
    \centering
    \includegraphics[width=\columnwidth]{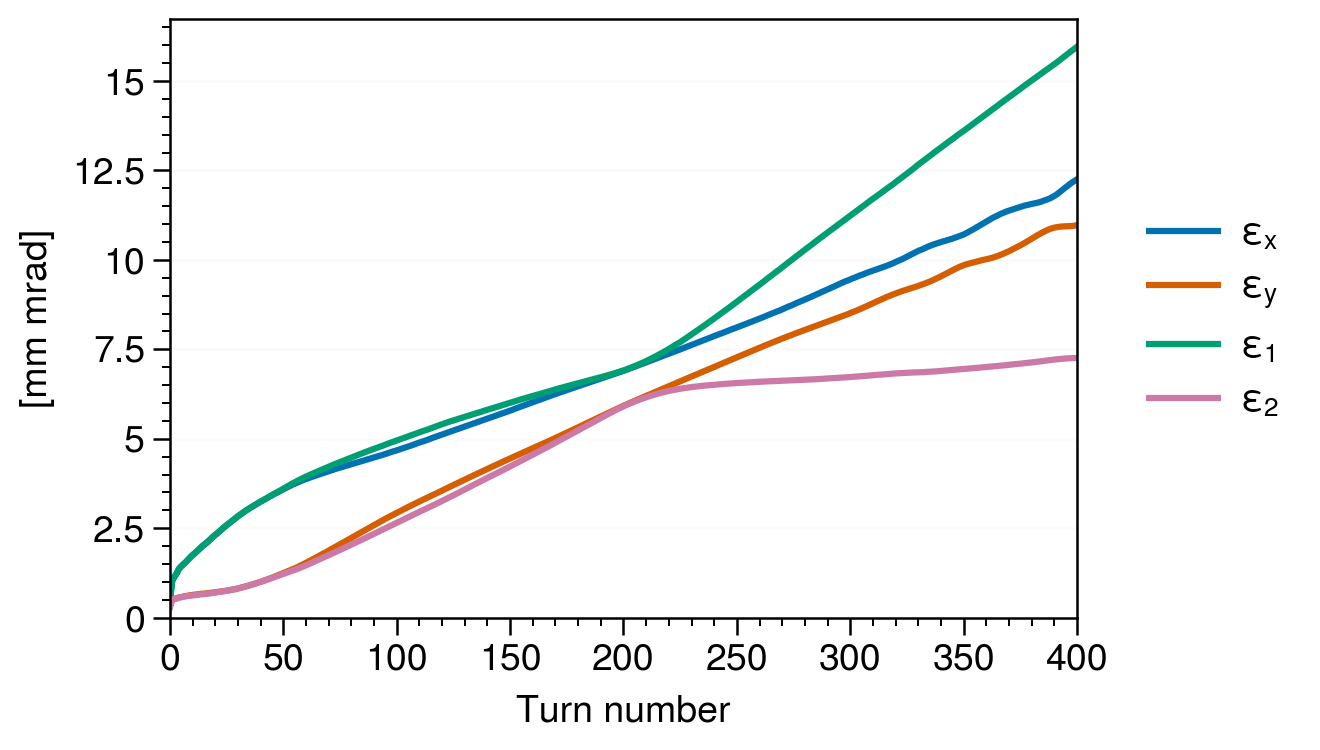}
    \caption{Simulated emittance evolution of with ring/injection parameters similar to those in the measurement in Fig.~\ref{fig:7}d.}
    \label{fig:8}
\end{figure}

As mentioned at the beginning of this section, a detailed discussion of the measured beam evolution is beyond the scope of this paper. Our main emphasis is that the data represented in Fig.~\ref{fig:7} was collected in approximately four hours, fitting within two rarely-occurring beam study periods with long machine setup times, and allows qualitative comparisons with computer simulation which can be used to guide future experiments. One example of such a comparison is found in Fig.~\ref{fig:8}, which shows a simulation of injection into the SNS ring with similar ring/injection parameters as Fig.~\ref{fig:7}d. 

\section{Conclusion}
\label{sec:Conclusion}

One critical component of efforts to generate a beam with small 4D emittance in the Spallation Neutron Source (SNS) is to measure the 4D emittance throughout accumulation. A variant of the multi-optics emittance reconstruction method was implemented using four available wire-scanners in the ring-target-beam-transport (RTBT) section of the SNS. The fixed-optics reconstruction method was not possible with the nominal machine optics: the probability of failed fits was large, and there was a significant uncertainty/bias in intrinsic emittances in the successful fits. Monte Carlo simulations were employed to select a new set of quadrupole strengths, eliminating the problem of failed fits and reducing the bias and uncertainty to acceptable values. 

The effect of beam mismatch on the reconstruction accuracy was also studied by varying the initial beam parameters and repeating the Monte Carlo trials. This increased the bias in the reconstructed intrinsic emittances from 7\% to 12\%, making it difficult to measure small cross-plane correlation in the beam; however, this bias decreased when significant cross-plane correlation was present in the beam. Thus, we concluded that the fixed-optics method is sufficient for faster (albeit less accurate) measurements of the intrinsic beam emittances in the SNS.

Finally, the usefulness of the fixed-optics method was demonstrated by reconstructing the emittance evolution during accumulation in the SNS ring. As expected, standard SNS painting produced very little cross-plane correlation. Initial tests of "elliptical painting" revealed noticeable changes in the 4D emittance evolution in qualitative agreement with simulations with similar ring/injection parameters. These results will be used to guide future experiments at the SNS.

\section{Acknowledgements}
This manuscript has been authored by UT-Battelle, LLC, under Contract No. DE-AC05-00OR22725 with the U.S. Department of Energy. The United States Government retains, and the publisher, by accepting the article for publication, acknowledges that the United States Government retains a nonexclusive, paid-up, irrevocable, world-wide license to publish or reproduce the published form of this manuscript, or allow others to do so, for United States Government purposes.

\appendix
\section{Clarification of experimental results}

\begin{figure*}[!t]
    \centering
    \begin{subfigure}{0.48\textwidth}
        \includegraphics[width=\linewidth]{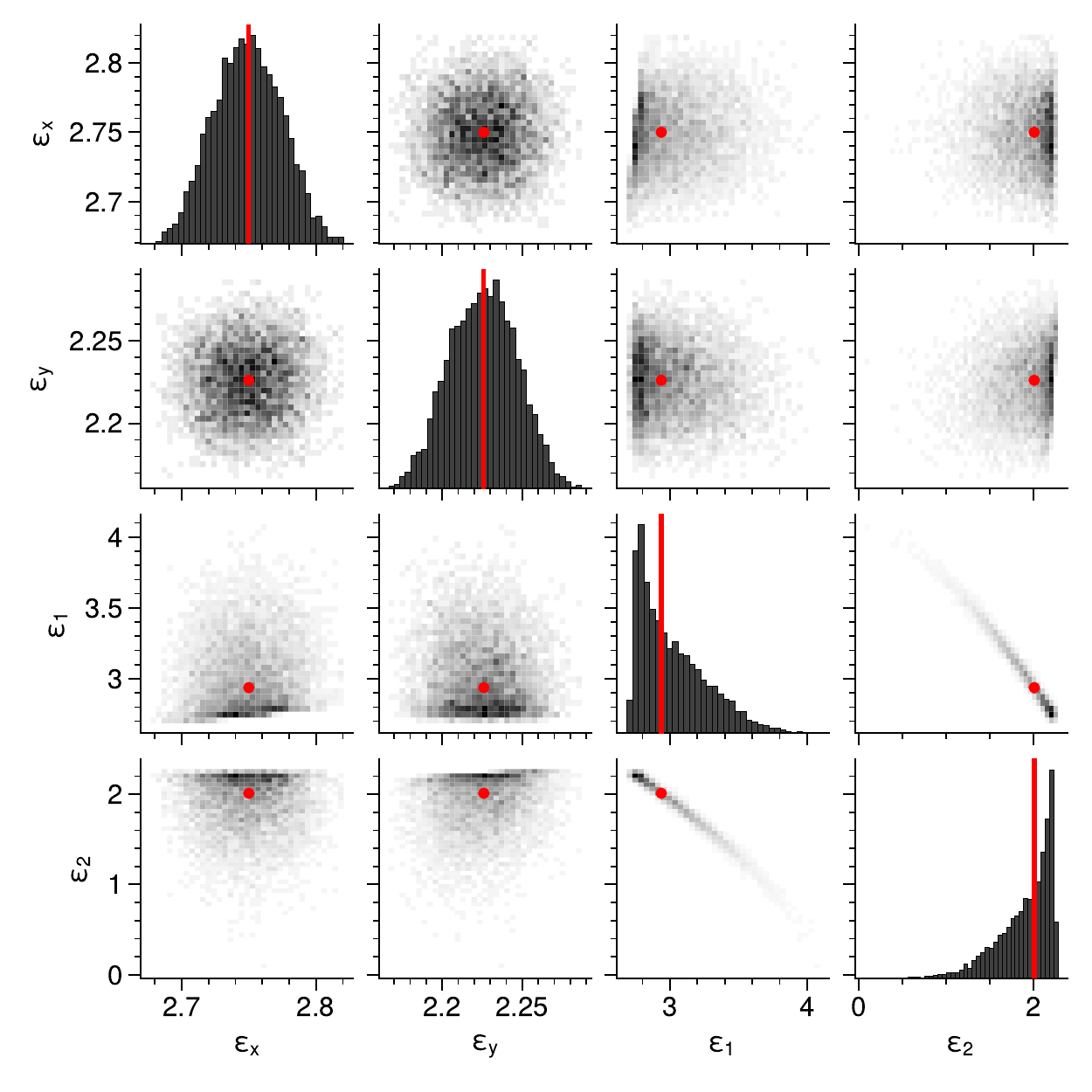}
        \caption{}
        \label{fig:9a} 
    \end{subfigure}
    \hspace{0.4cm}
    \begin{subfigure}{0.48\textwidth}
        \includegraphics[width=\linewidth]{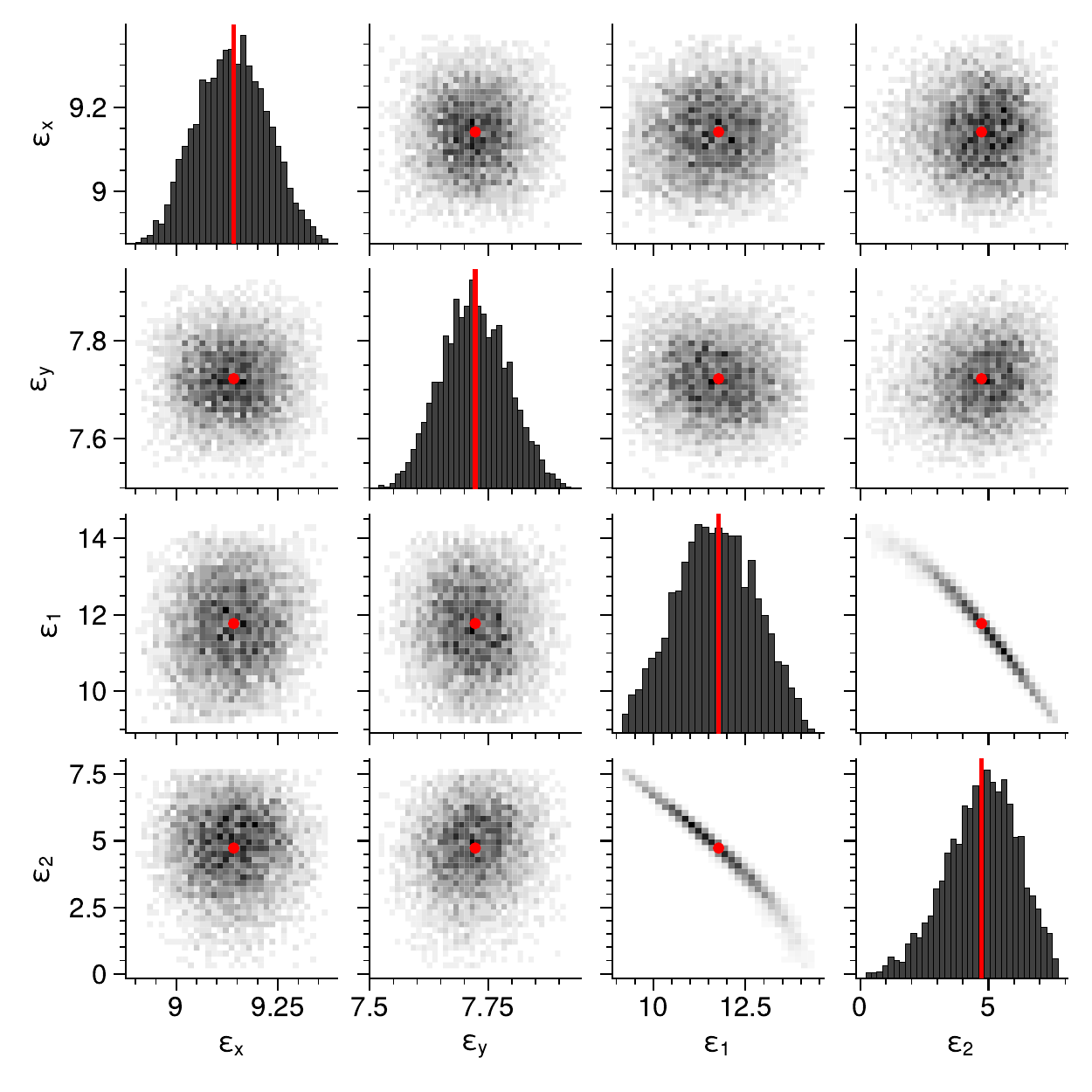}
        \caption{}
        \label{fig:9b}
    \end{subfigure}
    \caption{Emittance distribution used to produce the error bars in Fig.~\ref{fig:7}d at (a) 150 turns (b) 400 turns. 5000 reconstructions were performed from random 3\% perturbations of the measured moments. The measured emittances are plotted in red.}
    \label{fig:9}
\end{figure*}
\begin{figure}[t!]
    \centering
    \includegraphics[width=\columnwidth]{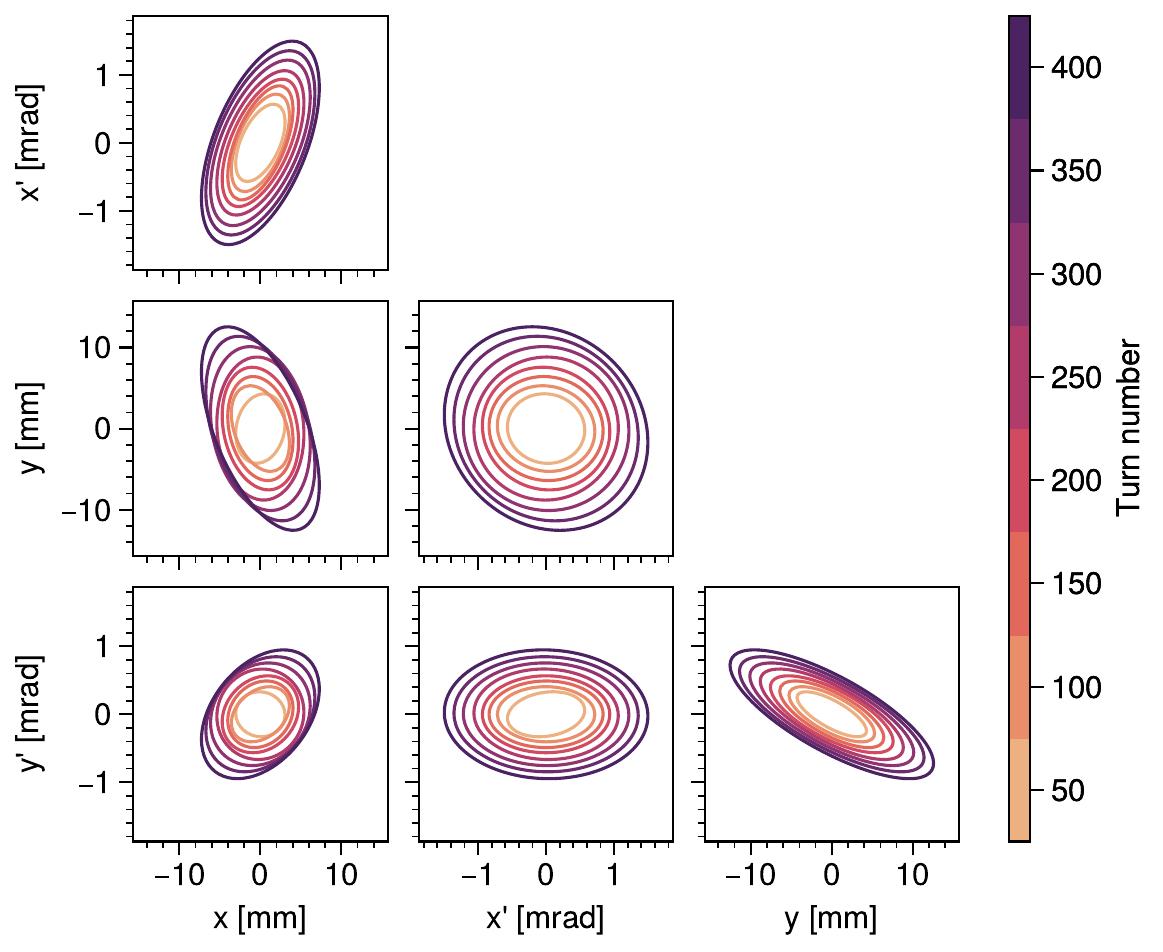}    
    \caption{2D projections of the reconstructed covariance ellipsoid as a function of turn number during injection in the SNS ring. The measured emittances are shown in Fig.~\ref{fig:7}d.}
    \label{fig:10}
\end{figure}
In this appendix, we offer two clarifications on the measurements in Fig.~\ref{fig:7}d. First, Fig.~\ref{fig:9} displays the distribution of apparent and intrinsic emittances used to generate the error bars in Fig.~\ref{fig:7}d at turn 150 and turn 400. Again, these emittances were obtained by repeating the reconstruction as 3\% noise was added to the measured $\langle{xx}\rangle$, $\langle{yy}\rangle$ and $\langle{xy}\rangle$. Notice that in (a), when the cross-plane correlation is small, the mean of the distribution is not exactly equal to the measured emittances, leading to the asymmetric error bars in Fig.~\ref{fig:7}d.

Second, since the intrinsic emittances are complicated functions of the second-order beam moments, it is not always obvious how changes in the intrinsic emittances affect the beam distribution. Therefore, in Fig.~\ref{fig:10}, we have plotted the 2D projections of the covariance ellipsoid for each measurement in Fig.~\ref{fig:7}d. The change in intrinsic emittances manifests in a tilting of the ellipses in planes other than $x$-$x'$ and $y$-$y'$.

\bibliographystyle{elsarticle-num} 
\bibliography{main.bib}





\end{document}